\documentclass[]{pasj02}

\usepackage[switch,mathlines]{lineno} 
\usepackage{url}
\usepackage{natbib} 

\jyear{2024}
\Received{}
\Accepted{}

\DeclareRobustCommand{\ion}[2]{\textup{#1\,\textup{#2}}}

\begin{document} 

\title{The extreme starburst J1044+0353 blows kpc-scale bubbles}

\author{
Edmund Christian \textsc{Herenz}\,\altaffilmark{1}$^{,\dag}$\altemailmark\orcid{0000-0002-8505-4678}\email{edmund.herenz@iucaa.in},
  Haruka \textsc{Kusakabe}\,\altaffilmark{2}$^{,}$\altaffilmark{3}\altemailmark\orcid{0000-0002-3801-434X}\email{haruka.kusakabe.takeishi@gmail.com},
  Soumil \textsc{Maulick}\,\altaffilmark{1}\orcid{0009-0003-8568-4850}
}

\altaffiltext{1}{Inter-University Centre for Astronomy and Astrophysics (IUCAA), Pune
  University Campus, Pune 411 007, India}
\altaffiltext{2}{National Astronomical Observatory of Japan, 2-21-1 Osawa, Mitaka, Tokyo
  181-8588, Japan}
\altaffiltext{3}{Department of General Systems Studies, Graduate School of Arts
  and Sciences, The University of Tokyo, 3-8-1 Komaba, Meguro-ku, Tokyo, 53-8902, Japan}


\maketitle

\begin{abstract}
  We report the discovery of a kpc-scale lacy filamentary structure in low surface-brightness H$\alpha$ and [\ion{O}{III}]$\lambda5007$ emission around the low-redshift, extremely metal-poor and compact reionisation-era analogue SDSS J1044+0353.  We identify seven elliptical arcs in H$\alpha$ emission at $\mathrm{SB}_\mathrm{H\alpha} \sim 1 - 2 \times 10^{-18}$erg\,s$^{-1}$cm$^{-2}$arcsec$^{-2}$.  We interpret these features as limb-brightened giant-shells that bound egg-shaped super bubbles with 3-3.5\,kpc radii.  These shells are significantly larger than the known giant shells around nearby star-forming dwarfs.  Kinematic maps reveal a gradient perpendicular to the major-axis and line broadening in the outskirts.  The latter, when interpreted due to line-splitting of the expanding shells, suggests expansion velocities of $\sim 40$\,km\,s$^{-1}$.  Notably, the properties of the giant shells evade description by simple analytic prescriptions for energy- and momentum-driven super-bubbles.
\end{abstract}


\defcitealias{Peng2023}{P23}
\defcitealias{Martin2024}{M24}


\footnotetext[$\dag$]{Vaidya-Raychaudhuri Fellow}

\section{Introduction}
\label{sec:intro}

Winds on galaxy scales driven by feedback from young stellar populations are key for understanding galaxy evolution and cosmology.  Outflows from starbursting low-mass ``dwarf'' galaxies are of special interest.  Such galaxies are characterised by spatially and temporally strongly clustered star-formation that leads to a concentrated injection of energy and momentum into their interstellar medium (ISM).  Conjoining with their shallow gravitational potential, this should efficiently produce large scale winds that launch material far beyond their main stellar bodies, especially if multiple bursts occur on timescales shorter than the dynamical time of the galaxy \citep[see review by][]{Collins2022}.  Moreover, these so called outflows are expected to be the main mechanism for creating channels of low neutral hydrogen columns in the circum-galactic gas through which hydrogen ionising photons can leak into the intergalactic medium.  In the early Universe this induces the phase transition from a neutral to an ionised intergalactic medium during the so-called Epoch of Reionisation \citep[see review by][]{Erb2015}.  Detailed observational inferences regarding the astrophysics of feedback-driven winds and outflows of relevance during the Epoch of Reionisation can be obtained by studying low-redshift analogues of such early galaxies, i.e. low-mass star-forming galaxies that are extremely metal poor and compact.  But, conventional spectroscopy, even with sophisticated analysis techniques \citep[e.g.][]{Xu2023,Amorin2024}, can only provide limited information regarding the spatial and kinematic distribution of the material entrained in the wind.

Generally, the ionised ``warm'' $T_e\sim10^4$ phase of the wind can be mapped in line emission, with H$\alpha$ and [\ion{O}{III}]$\lambda5007$ being the most luminous tracers.  Still, this phase shines only at surface luminosities (SL) several orders of magnitude fainter than emission from the ISM.  The detectability of this outflowing material is thus greatly enhanced if the galaxy scale wind expels material perpendicular to the line of sight.  Even so, classical narrow-band imaging, being limited in sensitivity and contrast, allowed to map such extraplanar outflows only for nearby ($\lesssim10$\,Mpc) systems \citep[e.g.][]{Marlowe1995,CMartin1998,Lee2016,McQuinn2019}. However, within this volume compelling early universe analogues are almost absent\footnote{The extremely metal-poor galaxy I Zw 18 is a famed exception \citep[e.g.][and references therein]{Arroyo-Polonio2024}.}.  Higher contrast and sensitivity can be achieved with image-slicer based integral field spectrographs on 8\,m class telescopes, namely the Keck Cosmic Web Imager (KCWI) and the Multi Unit Spectroscopic Explorer (MUSE).  This observational technique has now revealed a handful of spatially resolved maps of spectacular kpc-scale winds from extreme star-bursting dwarf galaxies at larger distances \citep{Bik2018,Menacho2019,Herenz2023b,McPherson2023}.

Here we present another such discovery for the outstanding nearby early universe analogue SDSS J1044+0353 (hereafter J1044). This low-redshift\footnote{For reference, the spectroscopic redshift of J1044, $z=0.012873$ \citep{Shim2013}, corresponds to a luminosity distance of $D_\mathrm{L}=55.7$\,Mpc and an angular scale of 0.278 kpc/\arcsec{} for there here assumed $(H_0,\Omega_M,\Omega_\Lambda)=(70\,\mathrm{km\,s}^{-1}\mathrm{Mpc}^{-1}, 0.3, 0.7)$ concordance cosmology.} blue compact galaxy was one of the first additions from the Sloan Digital Sky Survey to the limited pre-SDSS zoo of extremely metal poor galaxies at low-$z$ (\citealt{Kniazev2003}, cf. review \citealt{Kunth2000}).  As such it attracted significant interest in numerous sample-based observational follow-up studies that statistically investigate the physics of galaxy formation in extreme environments \citep[e.g.][]{Papaderos2008,Izotov2012,Nakajima2022,Isobe2023,Xu2024,Hatano2024}.  J1044 sticks out from those samples as a genuinely extreme emission line galaxy.  The galaxy is of low-mass ($\log_{10}(M/\mathrm{M}_\odot) \simeq 6.5$), extremely metal poor ($Z/Z_\odot = 0.058$), young ($\lesssim$20\,Myr), and highly star-forming ($\mathrm{SFR} \approx 0.2$\,M$_\odot$yr$^{-1}$; mass doubling time $\approx 4.4\times10^7$yr).  Comparable optical and ultraviolet spectral characteristics are found rarely at low $z$, but occur frequently amongst young star-forming galaxies in the early universe \citep{Berg2016}.  The spectroscopic similarity to high-$z$ galaxies is especially strengthened by the presence of a Balmer and Paschen jump due to strong nebular continuum emission from the extreme starburst.  This unique nature of the target motivated further deep optical Large Binocular Telescope MODS spectroscopic \citep{Berg2021,Olivier2022} and KCWI integral-field spectroscopic follow-up programmes \citep[][hereafter P23, M24]{Peng2023,Martin2024}.  These studies highlight the presence of a strong ionising radiation field that is incommensurate with even the most extreme stellar population synthesis models, but also incompatible with an AGN.  They also reveal the complex chemo- and thermodynamics of the ISM caused by intense feedback from the starburst.  Especially, \citetalias{Martin2024} provide significant clues that hint at the presence of large scale wind features.  However, the limited field of view of the KCWI observations (20.4\arcsec{}$\times$16\arcsec{}) did not allow to map the full extent of the large scale wind or outflow.  Here we analyse archival ESO/VLT MUSE integral field spectroscopic observations of J1044 that unveil the spectacular structure of the large scale feedback.


\section{Data}
\label{sec:data}

We obtained publicly available ESO/MUSE data for J1044 from the ESO archive.  These laser-assisted adaptive optics (AO) observations were taken in the wide field mode extended configuration of MUSE on UT dates 29/12/2019 and 30/12/2019.  The total exposure time is 4864\,s, split over eight exposures. No sky-offset frames were taken.  An inspection of the datacube reduced by ESO\footnote{\url{https://archive.eso.org/dataset/ADP.2021-05-17T13:24:16.108}} reveals spatially extended emission in H$\alpha$ ($\lambda_\mathrm{obs} = 6647.3$\,\AA{}) and [\ion{O}{III}]$\lambda5007$ ($\lambda_\mathrm{obs} = 5071.3$\,\AA{}).  However, this datacube also shows strong negative residuals at these wavelengths.  As detailed in \cite{Herenz2023b}, faint extended line emission filling the MUSE field of view (FoV) tends to be over-subtracted by the sky-subtraction procedure of the MUSE pipeline, since the pipeline determines the sky spectrum from continuum faint spaxels within the FoV \citep{Weilbacher2020}.  To mitigate this over-subtraction we re-reduced the raw observational data, then we modified the pipeline determined sky spectrum by linear interpolation over the affected wavelength ranges, and finally we used this corrected sky spectrum as input in a second run of the pipeline \citep[see also][]{Wofford2021}.  Lastly, we registered the spatial world coordinate system of the datacube visually against publicly available PanSTARRS images \citep{Chambers2016}.  Our datacube covers the wavelength range from 4600\,\AA{} to 9350\,\AA{}, sampled at 1.25\,\AA{}, with a gap from 5755\,\AA{} to 6007\,\AA{} around the AO laser wavelength.  The square spaxels are sampled at $0.04$\,arcsec$^2$.  By fitting a 2D Gaussian to compact sources in the field, we obtain a FWHM of $\simeq0.65$\,\arcsec{} in a synthesised $R$-Band from the datacube as a measure of the image quality.


\section{Analysis \& Results}
\label{sec:res}

\begin{figure}[t!]                  
  \begin{center}
    \includegraphics[width=0.48\textwidth]{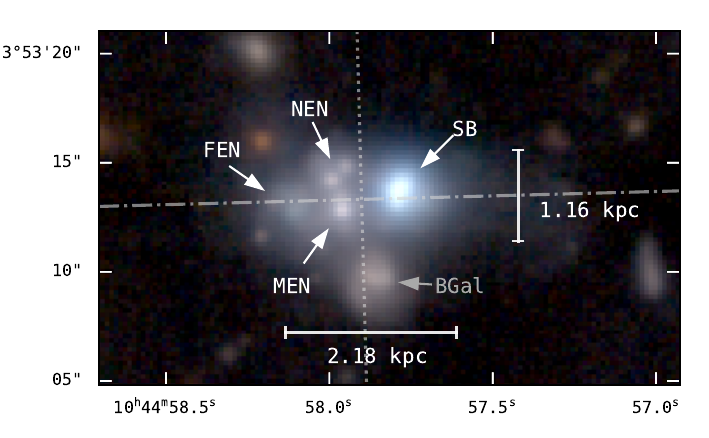}
  \end{center}
  \caption{False-colour image of J1044 using synthesised continuum images from the MUSE datacube (\emph{V}-, \emph{R}-, and \emph{I}-band are mapped to blue, red, and green, respectively).  The major- and minor axes are indicated by a dash-dotted and dotted line, respectively.  The shown field of view is 26.6\arcsec{}$\times$16.2\arcsec{} corresponding to 7.4\,kpc$\times$4.5\,kpc (see box in the left panel of Fig.~\ref{fig:haoiiinb}).  Regions are labelled according to the nomenclature of \citetalias{Peng2023}: SB\,$\rightarrow$\,main star burst, \{F,M,N\}EN\,$\rightarrow$\,\{faint, middle, north\}-eastern nucleus, whereas BGal labels a background galaxy that is very close in projection.  Other sources in this image are background galaxies as well.}
  \label{fig:cont}
\end{figure}

\begin{figure*}[t!]
  \begin{center}
    \includegraphics[width=0.95\textwidth,trim=0 10 3 3,clip=true]{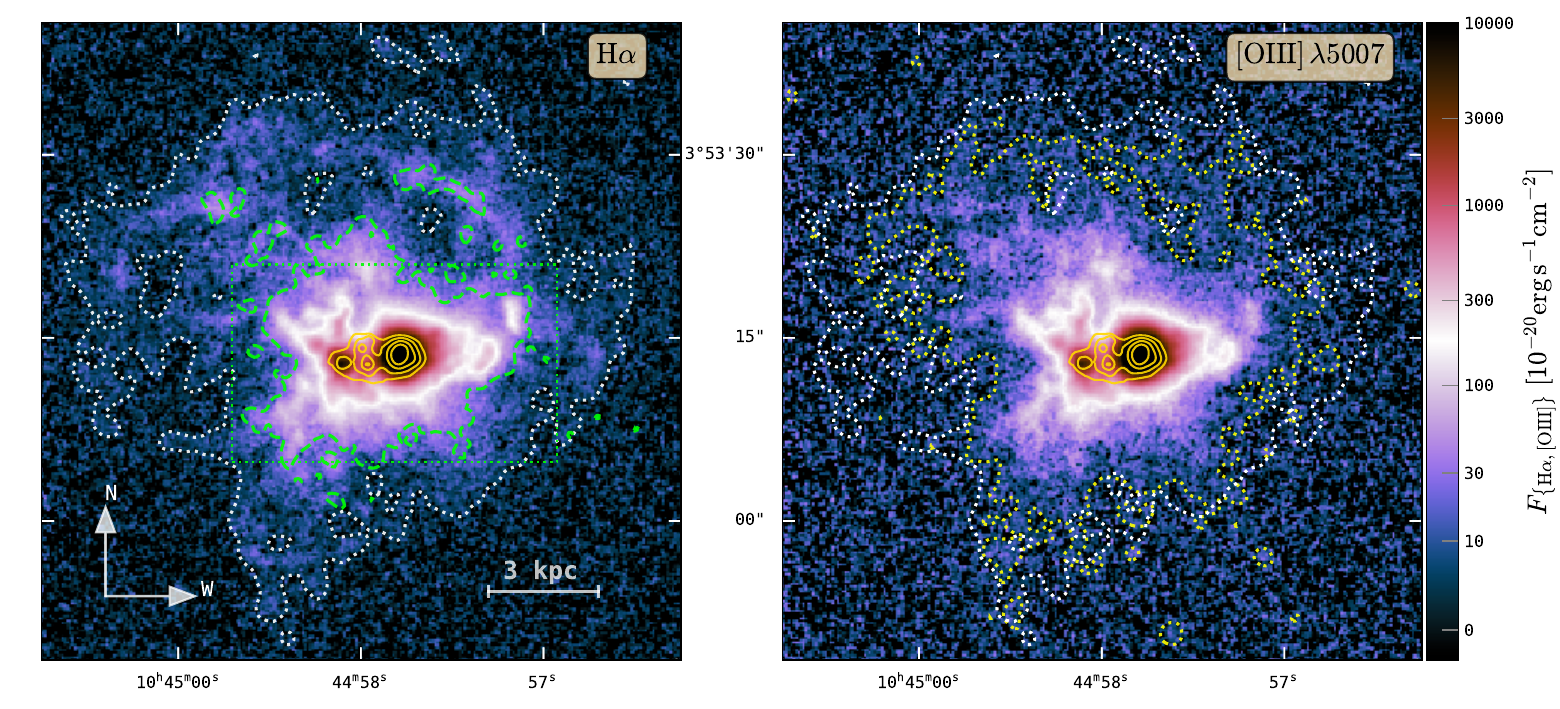} 
  \end{center}
  \caption{Optimally extracted H$\alpha$ (\emph{left}) and [\ion{O}{III}]$\lambda5007$ (\emph{right}) image of J1044.  Flux values, $F_\mathrm{\{H\alpha,[OIII]\}}$, are displayed with a logarithmic stretch, $\ln(3000 \times F)/\ln(3000)$, from $-1.5\times10^{-20}$ to $10^{-16}$erg\,s$^{-1}$cm$^{-2}$ as indicated by the colour-bar on the right; tick marks correspond SB values of $\mathrm{SB}_\mathrm{\{H\alpha,[OIII]\}} = \{2.5, 7.5, 25, 75, 250, 750, 2500 \}\times 10^{-18}$erg\,s$^{-1}$cm$^{-2}$arcsec$^{-1}$.  We also show contours that outline the detection limit in H$\alpha$ (white dotted lines in both panels), H$\beta$ (green dashed line in the H$\alpha$ panel), and [\ion{O}{III}] (yellow dotted line in the [\ion{O}{III}] panel).  Solid gold contours in the centre of both panels indicate the morphology of the continuum from the synthesised $R$-Band.  The rectangular field of view (FoV) of the false-colour continuum image of Fig.~\ref{fig:cont} is demarcated by green dashed line in the H$\alpha$ image.  The displayed field of view in both panels is 52.4\arcsec{}$\times$52.2\arcsec{} (14.57\,kpc$\times$14.51\,kpc).}
  \label{fig:haoiiinb}
\end{figure*}

\begin{figure}[b!]
  \begin{center}
    \includegraphics[width=0.5\textwidth]{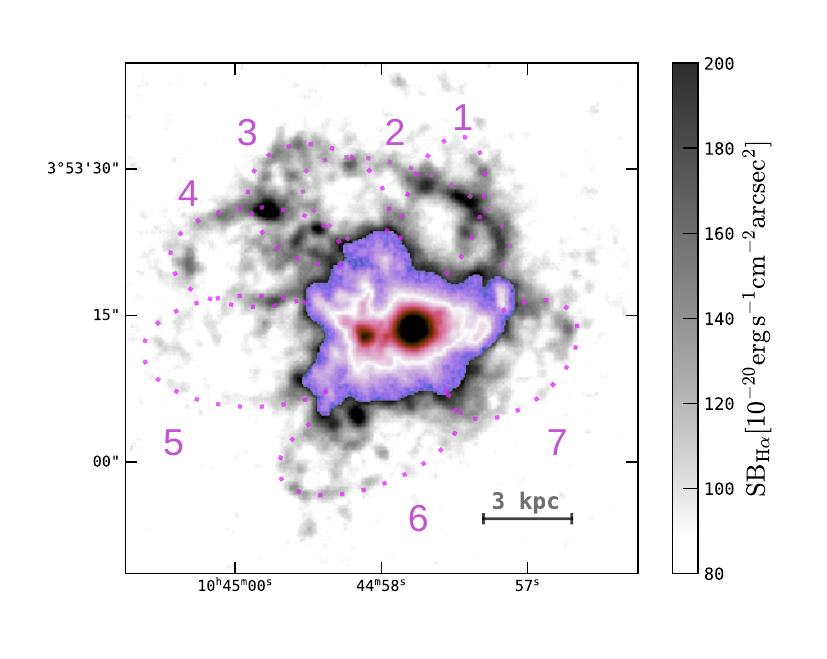} 
  \end{center}
  \vspace{-1.4em}
  \caption{Enhanced visualisation of filamentary low-SB H$\alpha$ emission (see Sect.~\ref{sec:res}).  In the higher-SB region ($\mathrm{SB}_\mathrm{H\alpha} > 2\times10^{-18}\,\mathrm{erg\,s^{-1}s^{-2}arcsec^{-2}}$) we inset the H$\alpha$ NB from Fig.~\ref{fig:haoiiinb}.  Elliptical arcs are drawn with magenta dotted lines to indicate the seven prominent kpc-scale filaments.  The displayed FoV is identical to the panels shown in Fig.~\ref{fig:haoiiinb}.}
  \label{fig:arcs}
\end{figure}

\begin{figure}[b!]
  \begin{center}
    \includegraphics[width=0.4\textwidth]{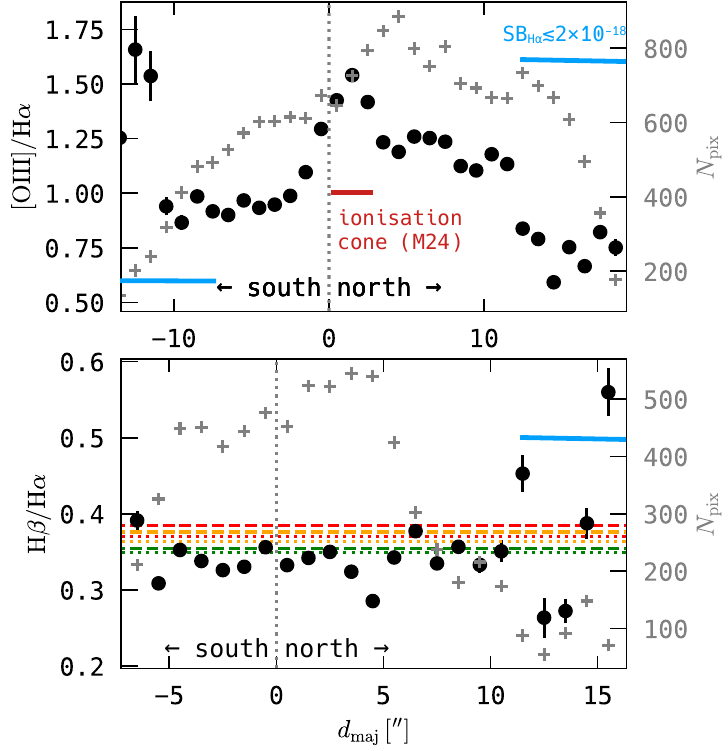}
  \end{center}
  \caption{Vertical line-ratio profiles, [\ion{O}{III}]/H$\alpha$ (\emph{top}) and H$\beta$/H$\alpha$ (\emph{bottom}), measured in 1\arcsec{} wide slits parallel to the major axis, with $d_\mathrm{maj}$ being the distance of the slit centre to the major axis.  Filled black circles indicate the galactic foreground extinction corrected ratios within 1\arcsec{} wide bins.  Grey crosses indicate the number of spaxels contributing to each vertical bin (labelled on the right ordinate).  The horizontal dashed (dotted) green, orange, and red line in the bottom panel show the theoretical case-A (case-B) recombination H$\beta$/H$\alpha$ ratios for $T=\{1,2,3\}\times10^4$\,K, respectively \citep{Storey1995}.  Thick cyan lines indicate the low-SB region where $\mathrm{SB}_\mathrm{H\alpha} < 2\times10^{-18}\,\mathrm{erg\,s^{-1}s^{-2}arcsec^{-2}}$ (cf. Fig.~\ref{fig:arcs}) and the thick red lines in the top panel marks the region of the ionisation cone identified in \citetalias{Martin2024}.}
  \label{fig:lrprof}
\end{figure}

\begin{figure*}[t!]
  \begin{center}
    \includegraphics[width=0.95\textwidth]{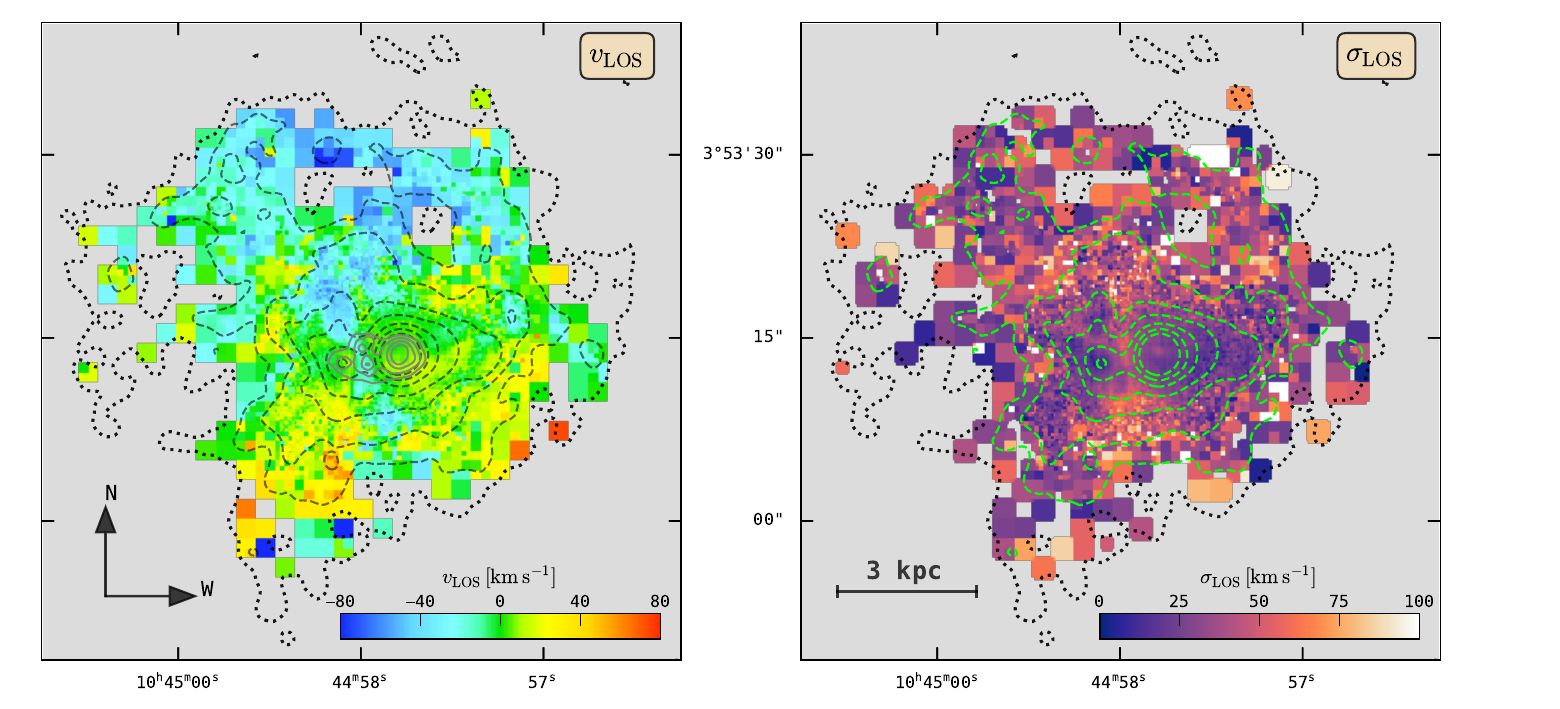}
  \end{center}
  \caption{Kinematic maps of J1044's galactic and CGM H$\alpha$ emission.  The \emph{left panel} shows the line-of-sight velocities, $v_\mathrm{LOS}$ in km\,s$^{-1}$ and the \emph{right panel} shows the velocity dispersion along the line of sight, $\sigma_\mathrm{LOS}$.  Dashed contours (grey in the left panel and green in right panel are iso-flux contours according to the ticks on the colour-bar in Fig.~\ref{fig:haoiiinb}, the solid contours indicate the morphology of the continuum, and the dotted contour delineates the limiting surface-brightness from Fig.~\ref{fig:haoiiinb}.  The displayed FoV is identical to Fig.~\ref{fig:haoiiinb}. }
  \label{fig:velfield}
\end{figure*}

\begin{figure}[t!]
  \begin{center}
    \includegraphics[width=0.45\textwidth]{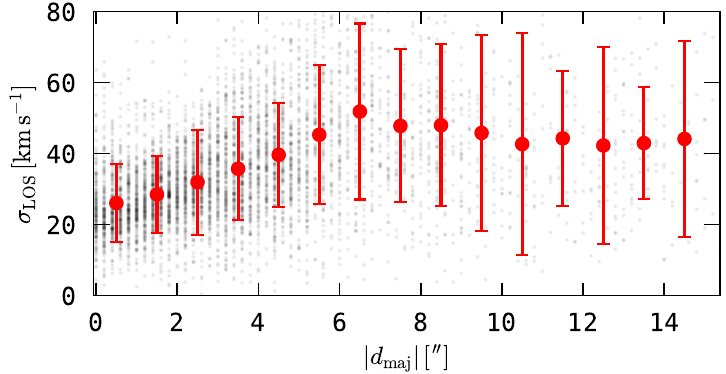}
  \end{center}
  \vspace{-0.7em}
  \caption{Velocity dispersion profile  measured perpendicular to the major axis.  Grey dots show individual spatial bins from the maps of Fig.~\ref{fig:velfield}, while the circles with error-bars are the mean and standard-deviation of those measurements in bins along $d_\mathrm{maj}$ of 1\arcsec{} width. } 
  \label{fig:prof}
\end{figure}

In Fig.~\ref{fig:cont} we present a synthesised false colour image of J1044 from the MUSE data.  This image highlights the multi-clump morphology and the heterogeneous spectro-photometric characteristics of this compact galaxy.  We label clump complexes according to \citetalias{Peng2023}, noting that the MUSE AO data resolves region NEN into two complexes.  HST imaging reveals that these each of these star-forming complexes consist of multiple individual clusters \citepalias[see Fig.~1 in][]{Martin2024}.

To set a frame of reference we define the centre of the galaxy and the major axis with the formalism of image moments \citep[e.g.][]{Stobie1980} applied to a binary mask obtained from thresholding the synthesised R-band image.  We choose a threshold of $1.7\times10^{-17}$erg\,s$^{-1}$cm$^{-2}$\AA{}$^{-1}$arcsec$^{-2}$, as below this limit no continuum is detected in the R-band; background galaxies are masked.  We obtain RA\,=\,\timeform{10h44m57.9s} , Dec\,=\,\timeform{+03D53'13.31''}, and $\theta_\mathrm{maj} = 1.54^\circ$ east of north.  As can be seen in Fig.~\ref{fig:cont}, this centre falls right between the regions NEN, MEN and SB.

For detecting and measuring extended line emission we removed the continuum by subtracting the output of 187.5\,\AA{} broad running median filter in each spaxel.  We then used the 3D filtering algorithm of \texttt{LSDCat} \citep{Herenz2023}.  Here we opted for a 3D Gaussian filter, whose spatial and spectral FWHM is 0.75\arcsec{} and 150\,km\,s$^{-1}$, respectively.  By visual inspection we found that this parameter combination optimally enhances the detectability of the low surface-brightness filamentary line emission without degrading the lacy morphology.  We detect line emission of an extent multiple times that of the main stellar body in H$\alpha$ and [\ion{O}{III}] as can be seen in Fig.~\ref{fig:haoiiinb} where we show optimally extracted NB images in both lines.  The optimal NB extraction procedure is described in Sect.~5.2.3 of \cite{Herenz2020}, and here we used a threshold of six in the \texttt{LSDCat} signal-to-noise cube.  This threshold corresponds to a limiting SB of $1.25\times10^{-18}$erg\,s$^{-1}$cm$^{-2}$arcsec$^{-2}$ in H$\alpha$ and $2.5\times10^{-18}$erg\,s$^{-1}$cm$^{-2}$arcsec$^{-2}$ in [\ion{O}{III}].  After correcting for foreground MW absorption\footnote{We use the \cite{Cardelli1989} extinction law and $E(B-V)_\mathrm{MW} = 0.037$ from \cite{Schlafly2011}. \label{fn:1}} this corresponds to limiting SLs of $7.34\times10^{36}$erg\,s$^{-1}$kpc$^{-2}$ in H$\alpha$ and $1.52\times10^{37}$erg\,s$^{-1}$kpc$^{-2}$ in [\ion{O}{III}].  We also detect extended H$\beta$ emission at a similar limiting SB / SL as [\ion{O}{III}], albeit only very fragmentary at larger extends, and not as extended as H$\alpha$ or [\ion{O}{III}].  We delimitate the limiting SBs / SLs by contour lines in Fig.~\ref{fig:haoiiinb}.

The NB images reveal multiple bright wisps and spurs emanating from the star-forming complexes.  Especially, the X-shaped structure centred on the FEN region appears noteworthy.  Such morphology occurs naturally as the limb brightened walls of a bi-conical outflow \citep[e.g.][]{Herenz2023b}.  On larger scales we observe a complex network of filamentary and lacy arcs that span over regions that are of similar or even larger size than the galaxy's main stellar body.  The outermost parts of these filaments, some of which appear seemingly unrelated to the features seen closer to the star-forming complexes, are not as pronounced in [\ion{O}{III}], but this is also related to the increased background noise in the blue spectral range of the datacube.  Parallel to the major and minor axes we measure a projected maximum extent of 12.5\,kpc (9.7\,kpc) and 11.6\,kpc (10.2\,kpc) in H$\alpha$ ([\ion{O}{III}]), respectively.  The contribution from the extended H$\alpha$ ([\ion{O}{III}]) emission outside the main stellar body (delineated by the outermost gold contours in Fig.~\ref{fig:haoiiinb}) to the total emission line flux ($F_\mathrm{H\alpha} = 1.4\times10^{-13}$erg\,s$^{-1}$cm$^{-2}$ / $F_\mathrm{[OIII]} = 1.9\times10^{-13}$erg\,s$^{-1}$cm$^{-2}$) is only 10\% (7\%).

To better grasp the morphology of the low-SB filaments we smooth the optimally extracted H$\alpha$ NB with a square top-hat filter of 1.2\arcsec{} width.  We show this improved visualisation in Fig.~\ref{fig:arcs}, where the contrast is tuned to enhance the narrow SB-range at which these filaments shine.  Additionally, we inset in Fig.~\ref{fig:arcs} the H$\alpha$ visualisation from Fig.~\ref{fig:haoiiinb} in the region with $\mathrm{SB}_\mathrm{H\alpha} > 2 \times 10^{-18}$erg\,s$^{-1}$cm$^{-2}$arcsec$^{-2}$.  The contribution of the filamentary H$\alpha$ flux outside this region to the total H$\alpha$ flux is merely 1.5\%.  In Fig.~\ref{fig:arcs} we perceive that most of the filaments can be reconciled with elliptical arcs whose start and end points connect to the spurs seen at higher fluxes.  To convey this point, we draw in this figure seven elliptical arcs (labelled 1 to 7).  Guided by the polar geometry of large-scale H$\alpha$ shells in nearby star-forming galaxies \citep[e.g.][]{CMartin1998,Sanchez-Cruces2015} we found this as the best fitting configuration by manual placement and alignment.  While this simple geometric configuration does not capture the full complexity of the filamentary web, it is a salient approximation of the most prominent features.  For arcs 2, 6 and 7 (1, 3, and 5) the minor (major) axis intersects with the main stellar body, but such an alignment can not be established for arc 4.  Arcs 5, 6, and 7 show large gaps pointing away from the galaxy (``bull horn''-like morphology), whereas the other arcs exhibit fragmentation on smaller scales.  The radii of the major axes of the ellipses that span the arcs are found to be in between 3 to 3.5\,kpc.  Large-scale H$\alpha$ emitting arcs surrounding star-forming complexes in nearby star-bursts are naturally explained as limb-brightened giant-shells that surround outflow driven super-bubbles \citep[e.g.][]{Marlowe1995,CMartin1998}.  The largest such shells around nearby dwarf galaxies are about one kpc in diameter and surround I\,Zw\,18 \citep{Martin1996} and NGC\,1569 \citep[][]{Sanchez-Cruces2015,Hamel-Bravo2024}.  In analogy to those systems we regard the found filaments also as such limb-brightened shells.  Notably, the shells surrounding J1044 are more than five to eight times larger than the largest shells known around nearby dwarf galaxies.

The three emission lines in which we detect extended emission allow for limited insights into the physical state of the outflowing gas.  To this aim we show in Fig.~\ref{fig:lrprof} line-ratio profiles\footnote{Corrected for galactic extinction; see footnote~\ref{fn:1}.} of [\ion{O}{III}]/H$\alpha$ and H$\beta$/H$\alpha$ \citep[cf.][Sect.~3.2]{Herenz2023b}.  We extracted these profiles in 1\arcsec{} wide slits that are oriented parallel to the major axis.  The flux integration region in each slit is defined by the detection contour of the weaker line in the nominator and the vertical profile is terminated when there are less than 50 spaxels within this boundary.  The resulting H$\beta$/H$\alpha$-profile is mostly flat and compatible with Balmer-decrements under classical Case-A or Case-B recombination assumptions (i.e. $\tau_{\mathrm{Ly}n} = 0$ or $\tau_{\mathrm{Ly}n} = \infty$, where $\tau_{\mathrm{Ly}n}$ is the optical depth of all Lyman-series lines).  Smaller ratios than Case-A/B may be explained by the presence of dust.  However, H$\beta$/H$\alpha$ ratios larger than Case-A/B (Balmer-decrement anomaly), as observed here in the northern low-SB region, may indicate recombination conditions intermediate between Case-A and Case-B \citep[dubbed Case-C, see][]{Ferland1999}.  There differences in optical depths of the Lyman-$\beta$ and Lyman-$\gamma$ transitions alter the H$\beta$/H$\alpha$ ratio \citep[Sect.~3.2 in][]{yanagisawa2024}.  While for very optically thin nebulae the resulting H$\beta$/H$\alpha$ decreases and thereby mimics the effect of dust reddening, it leads to the Balmer-decrement anomaly at higher neutral columns.  A more significant detection of this effect is possibly seen in the outflow of SBS\,0335-052E \citep{Herenz2023b}.  Alternative scenarios for the Balmer-decrement anomaly are proposed by \cite{yanagisawa2024} and \cite{Scarlata2024}, but they appear more suited for high-density interstellar media.  The [\ion{O}{III}]/H$\alpha$ ratios in the southern CGM are lower than in the north.  In the north the peak in the high-SB region appears related to the highly ionised cone mapped in [\ion{O}{III}]/[\ion{O}{II}] by \citetalias{Martin2024}.  In the low-surface brightness region towards the north, i.e. within the arc-like structures, [\ion{O}{III}]/H$\alpha$ decreases again, potentially indicating a less ionised medium that does not promote the escape of Lyman continuum photons.  The highest [\ion{O}{III}]/H$\alpha$ ratios are found at the southern tip of the extended emission, which is perhaps related to a high ionisation parameter or a hardening of the ionising spectrum due to Lyman continuum absorption.  We caution, however, that variations in [\ion{O}{III}]/H$\alpha$ are not only driven by the excitation of the plasma, as the ratio is also sensitive to temperature and metallicity.  Hence, the here proposed interpretations remain conjectural.

Lastly, we analyse the kinematics of the warm ionised phase by fitting the H$\alpha$ profiles with \texttt{lime} \citep{Fernandez2024} after spatially binning the data with \texttt{adabin} \citep{Li2023} to a target signal-to-noise ratio of 8 and ignoring bins larger than 4\,arcsec$^2$.  We found 1D Gaussians to be an adequate representation for the vast majority of the data (reduced $\chi^2 \lesssim 2$), especially in the CGM, except for spaxels near region SB, which require an additional broad line component (reduced $\chi^2 \gtrsim 30$), consistent with the findings of \citetalias{Martin2024}.  We correct the velocity dispersion measurements for line-spread-function (LSF) broadening using the LSF-model of \cite{Bacon2023}; at the observed H$\alpha$ wavelength the LSF dispersion is 48.4\,km\,s$^{-1}$.  We display the line-of-sight velocity and velocity dispersion maps from the single component Gaussian fit in Fig.~\ref{fig:velfield}.  Ionised gas kinematics of the ISM surrounding the star-forming complexes have been published previously (\citealt{Moiseev2010}; \citealt{Isobe2023}; \citetalias{Peng2023}) and are consistent with our result.  The newly revealed kinematics of the line emitting gas in the halo appear very complex.  There is a pronounced gradient from approaching to receding velocities perpendicular to the major axis (amplitude $\approx 45$\,km\,s$^{-1}$), which is typical for the presence of a large scale wind \citepalias{Peng2023}.  Nevertheless, the velocity field exhibits manifold discontinuities on smaller scales.  Interestingly, the blue-shifted perturbation seen on ISM scales \citepalias{Peng2023} extents to the halo.  Further outward in the halo discontinuities appear often correlated with the web of arcs.  They are likely the result of projection effects, i.e. the neighbouring sight-lines intersect different shells at different distances and orientations.  While the velocity dispersion in the ISM is relatively small ($\sigma_\mathrm{LOS} \approx 15$\,km\,s$^{-1}$), we observe an increase with distance from the minor axis (Fig.~\ref{fig:prof}).  The higher velocity dispersions in the halo ($\sigma_\mathrm{LOS} \approx 40 - 50$\,km\,s$^{-1}$) may indicate highly turbulent gas.  However, since the emission in the halo stems from multiple overlapping kpc-scale shells, sight-lines will frequently intersect front- and backside of these shells and the broadening may then be caused by the expansion of those shells \citep[cf.][]{Monreal-Ibero2023}.  Hence, at higher spectral resolution we would observe line splitting as observed in feedback blown super-bubbles of more nearby galaxies \citep[e.g.][]{Marlowe1995,CMartin1998,Ambrocio-Cruz2004,Sanchez-Cruces2015}.  Using a numerical experiment similar to the setup described in \cite{Robertson2013} we find that barely resolved lines split at $\Delta v \approx 75$\,km\,s$^{-1}$ would lead to an observed single line with LSF-corrected dispersion of 45\,km\,s$^{-1}$.  This allows us to approximate the expansion velocity of the shells as $v_\mathrm{exp} \leq \Delta v / 2 \approx 37.5$\,km\,s$^{-1}$, which is approximately half of (similar to) the expansion speeds measured for the kpc-scale bubbles in NGC\,1569 (I\,Zw\,18).

\vspace{-1em}

\section{Discussion \& Outlook}
\label{sec:disc}

We compare the observationally inferred properties of the uncovered
shells against classical analytical expectations for spherically symmetric expanding bubbles \citep[][and references therein]{Lancaster2021}.  The classical ``snowplough'' scenario assumes that radiative cooling is efficient only in the thin shell of shocked gas that surrounds the bubble.  This shock front is then energy-driven through an ambient medium of density $\rho$, with the energy inside the bubble being build-up by the mechanical wind luminosity $\mathcal{L}_w$ over time $t$.  For an energy-driven bubble the shell radius and velocity evolve as $R_\mathrm{E}=\left( ( 125 \mathcal{L}_w t^3 ) / (154\pi {\rho} ) \right )^{1/5}$ and $v_\mathrm{E} = \mathrm{d}R_\mathrm{E} / \mathrm{d}t = 3 R_\mathrm{E} / 5 t$, respectively.  For super-bubbles penetrating through the CGM the wind luminosity  $\mathcal{L}_w$ must be dominated by the ensemble of core-collapse supernovae (CCSNe) within the star-burst.  After a delay of $\Delta t \approx 3$\,Myr for the onset of CSSNe we expect an average $\mathcal{L}_w \simeq 3 \times 10^{40}$erg\,s$^{-1} (10^6$M$_\odot)^{-1}$ from the $Z=0.001$ starburst models of \cite{Leitherer1999}.  \citetalias{Peng2023} estimate ages of $t_\mathrm{SSP} \approx 15-20$\,Myr and a total mass of $M\simeq 4 \times10^6$M$_\odot$ for regions MEN, NEN, and FEN combined.  Compared to the massive and extremely young region SB ($3-4$\,Myr, $M=1.2\times10^6$M$_\odot$), where CCSNe have yet to occur, we assume these regions as the sources of $\mathcal{L}_w$ powering the bubbles.  Adopting $n_0 = \rho / (1.4 m_\mathrm{H}) = 0.1$\,cm$^{-3}$ as typical ambient hydrogen density of the CGM and setting $t=t_\mathrm{SSP} - \Delta t$ we find $R_\mathrm{E} \simeq 1.9 - 2.4$\,kpc and $v_\mathrm{E} \simeq 80 - 90$\,km\,s$^{-1}$.  These values are neither in agreement with the observed radii nor the estimated expansion velocity.  We thus consider the alternative scenario, where cooling beyond the shock front inside the bubble is extremely efficient.  In this case energy does not build up in the bubble, and so its evolution is solely determined by the momentum input rate $\dot{p} = \sqrt{2 \mathcal{L}_w \dot{M}}$, where $\dot{M}$ is the mass injection rate due to CCSNe, and the shell radius and velocity evolve as $R_\mathrm{p} = \left ((3 \dot{p} t^2)/(2 \pi \rho \right) )^{1/4}$ and $v_\mathrm{p} = \mathrm{d}R_\mathrm{p}/\mathrm{d}t = R_\mathrm{p} / 2t $.  From \cite{Leitherer1999} we obtain an average of $\dot{M}=2 \times 10^{-2}$M$_\odot$yr$^{-1} (10^6$M$_\odot)^{-1}$ after the onset of CCSNe, and with this we find $R=1.4-1.6$\,kpc and $v=47 - 56$\,km\,s$^{-1}$ for the momentum driven bubble.  Here the expansion velocities are in better agreement with the observations, but the radii are now significantly underpredicted.  Above considerations concern only a single bubble, but the discrepancies will become even more severe since the available $\mathcal{L}_w$ or $\dot{p}$ must be partitioned to blow seven super-bubbles.  If we assume equal partitioning $R_\mathrm{E}$ and $R_\mathrm{p}$ will decrease by a factor of $(1/7)^{1/5} = 0.68$ and $(1/7)^{1/4}=0.61$, respectively.  Thus, in contrast to $\lesssim 1$\,kpc super-bubbles around nearby starbursts \citep{Marlowe1995,CMartin1998}, the here found large-scale bubble structure around J1044 appears incommensurate with simple analytic prescriptions.

Morphologically the filamentary structure surrounding J1044 differs from the elongated linear filaments found in conical- or bi-conical winds that emanate from the nuclear star-clusters of more evolved galaxies such as, e.g., M\,82, NGC\,3079, NGC\,1482, or NGC\,4383 \citep[see][and references therein]{Watts2024}.  These structures extent up to tens of kpc from the stellar body and are of comparable size to the phenomenon observed here.  The wind driving star-forming nuclear star-super-clusters in these systems are embedded in evolved disk-like galaxies.  In contrast, metal-poor compact dwarfs are often gas-rich and embedded in kpc-scale HI envelopes.  Single dish HI observations indeed confirm that J1044 has a high gas-mass fraction and a long gas depletion time \citep[$f_\mathrm{HI} = 18.6$, $\tau_\mathrm{dep} = 1.5$\,Gyr;][]{Chandola2024}.  We thus suggest, that the formation of the kpc-scale shells is related to an extended HI envelope.  The shells then form as long as the wind sweeps up neutral gas from the halo.  Eventually, when they reach to halo regions of sufficiently low density, they will rupture and break due to Rayleigh-Taylor instabilities and thereby the hot gas and ionising radiation can leak further outwards.  Indeed, arcs 5, 6, and 7 appear broken, and are indicative that here such a bubble blow-out has happened.  Our findings thus reveal the mechanism by which a significant amount of hydrogen ionising photons can leak into the intergalactic medium.  In J1044 the bulk of the ionising photons are currently produced in region SB, while the mechanical luminosity from the slightly more evolved regions FEN, MEN, and NEN has transformed the originally optically thick CGM envelope into an environment that is transparent to ionising photons.  Thus, we likely resolve the effects two-stage/multi-stage starburst scenarios that have been suggested for ionising photon leaking galaxies (or candidates) from spectroscopic studies \citep[e.g.][]{Enders2023,Amorin2024}. 

The morphology and kinematics of J1044's outflow are also different from other large scale outflows from extreme starbursts uncovered with IFS: ESO\,338-IG\,04 \citep{Bik2018}, Haro\,11 \citep{Menacho2019}, Mrk\,1486 \citep{McPherson2023}, and SBS\,0335-052E \citep{Herenz2023b}.  The first three systems are more than two orders of magnitude more massive than J1044.  While the disk-like Mrk\,1468 exhibits a well defined large-scale bi-cone that extends more than 8\,kpc above and below the disk (in addition to outflowing material within the plane of the disk) the $\sim10$\,kpc outflows from irregular systems Haro\,11 and ESO\,338-IG\,04 appear morphologically and kinematically highly complex.  On the other hand, the main outflow from the system SBS\,0335-052E \citep{Herenz2023b}, which is more compact and of lower metallicity than J1044 but comparable in SFR and $M_*$, appears as a single $\sim 15$\,kpc cone emitting in narrow line emission.  Clearly defined arc-like shells on kpc-scales, as seen here, have not been found for any of those systems.  To understand what drives this observed diversity we need to increase the sample size of spatially resolved outflows from extreme starbursts.  In fact, as of yet it is even unknown whether large-scale low-SB ionised gas structures are a common feature of such galaxies.  The unprecedented sensitivity for low-SB line emission offered by IFS on 10\,m class telescopes has been demonstrated, e.g., by the detections of warm ionised CGM in even higher-redshift systems \citep{Zhang2024}, the discovery of 50\,kpc large outflow-bubbles around a massive $z\sim1$ active galaxy \citep{Rupke2019}, or the recent mappings outflows gas on $>10$\,kpc scales at $z\sim1$ in \ion{Mg}{II} emission \citep{Guo2023}.  Thus this observational technique appears ideally suited for such a census.

\vspace{-1em}

\begin{ack}
  We thank the anonymous referee for insightful comments that significantly improved the quality of this publication.  Based on observations made with ESO Telescopes at the La Silla Paranal Observatory under programme ID 0103.B-0531 (PI: D. Erb).
\end{ack}

\section*{Funding}
Haruka Kusakabe acknowledges support from the Japan Society for the Promotion of Science (JSPS) Research Fellowships for Young Scientists (202300224) and JSPS KAKENHI Grant Numbers JP23KJ2148 and JP25K17444.

\bibliographystyle{aasjournal}
\bibliography{J1044_outflow_paper.bib}

\end{document}